\documentclass[oldversion,letter]{aa}
\usepackage{times}
\usepackage{txfonts}
\usepackage{graphics}
\usepackage{xspace}
\usepackage{epsfig}
\usepackage{natbib}
\usepackage{rotating}

\newcommand{\lpsep}{LP\,412-31~}
\newcommand{\lp}{LP\,412-31}

\begin{document}

\title{Simultaneous optical and X-ray observations of a giant flare \\ on the ultracool dwarf LP\,412-31}

\author{B. Stelzer\inst{1} \and J. H. M. M. Schmitt\inst{2} \and G. Micela\inst{1} \and C. Liefke\inst{2}}

\institute{INAF - Osservatorio Astronomico di Palermo,
Piazza del Parlamento 1, I-90134 Palermo, Italy \and 
  Hamburger Sternwarte,
  Gojenbergsweg 112,
  D-21029 Hamburg, Germany}

\offprints{B. Stelzer}
\mail{B. Stelzer, stelzer@astropa.unipa.it}
\titlerunning{Giant optical and X-ray flare from the M8 dwarf LP\,412-31}

\date{Received $<$02-10-2006$>$ / Accepted $<$18-10-2006$>$} 

\abstract{
Cool stars are known to produce flares probably as a result of magnetic reconnection in their outer atmospheres. 
We present simultaneous {\em XMM-Newton} optical $V$ band and X-ray observations of the M8 
dwarf \lp. 
During the observation a giant flare occurred, with an optical amplitude of $6$\,mag and total energy 
of $3 \cdot 10^{32}$\,erg in each the $V$ band and soft X-rays. Both flare onset and flare decay were 
completely covered in both wavebands with a temporal resolution of $20$\,s, allowing for a determination 
of the flare time scales as well as a study of the temperature evolution of the flaring plasma.
The data are consistent with an impulsive energy release followed by radiative cooling
without further energy release during the decay phase. Our analysis suggests that the optical 
flare originates from a small fraction of the
surface of \lp, while the characteristic scale size of the flaring X-ray plasma is of the
order of the stellar radius or larger. The absence of any small-scale variability in the light curve 
suggests a non-standard flare number energy distribution.}
 

\maketitle

\section{Introduction}\label{sect:intro}

Flares, i.e., short-term enhancements of the emission in various wavebands, 
belong to the most obvious manifestations of solar and stellar magnetic activity.  
During a solar white-light flare the footpoints of magnetic loops light up in optical 
and ultraviolet emission lines and continuum. Intense X-ray emission goes along with such events, and 
is ascribed to the filling of these loops with heated plasma evaporated off the chromospheric layers along
the magnetic field lines into the corona
\citep[the `chromospheric evaporation scenario', e.g.,][]{Antonucci84.1}.  Depending on the flare frequency
distribution quiescent high-energy emission might be due to a
super-position of unresolved micro-flares \citep{Krucker98.1, Aschwanden00.1}.

On dM(e) stars (``flare stars'') flares are far more frequently observed (mostly in $U$ band or H$\alpha$) 
than on the Sun.  
However, because of significant logistic and observational difficulties,
simultaneous data taken in different wavebands, indispensable to test the validity of the above 
scenario, has rarely been secured. The situation becomes even worse for the low-mass end of
the main-sequence.  The fraction of H$\alpha$ emitters among the M dwarf population
decreases quickly for ultracool dwarfs beyond spectral type $\approx$ M7, 
yet objects of late-M or L spectral type 
can produce spectacular H$\alpha$ flares \citep{Liebert99.1, Liebert03.1, Fuhrmeister04.1}.
X-ray emission has been observed only from a rather small number of
ultracool dwarfs, and in most of these cases X-ray detections were made only during
flares \citep{Fleming00.1, Rutledge00.1, Schmitt02.1, Stelzer04.1}. 
None of these observations was accompanied by concurrent observations in other wavebands.
Optical broad-band observations of (ultracool) dwarfs are more abundant. However, most
of the available data do not have the temporal resolution required to follow the rapid evolution of the 
photospheric and chromospheric emission.  The observed magnitude increases in the optical wavebands
can be enormous.  Some -- rare -- examples of huge optical events on M dwarfs 
with amplitudes $>6$\,mag in optical and UV bands, respectively, are presented by
\citet{Pagano97.1} and \citet{Rockenfeller06.1}. 
Observational records of nine superflares in various wavebands were collected by 
\citet{Schaefer00.1}, but these observations lack any information on time evolution. 

In this letter we present a giant X-ray flare on the M8 dwarf \lp,
recorded with the {\em XMM-Newton} satellite with a time resolution of a few seconds
simultaneously in soft X-rays and in the $V$ band.
\lpsep was originally identified in a proper motion survey of \citet{Luyten79.1} and
first classified as a very late M dwarf by \citet{Kirkpatrick95.1}. 
Since then it has frequently been used as  
reference star for spectral type M8, although little is actually known about its stellar parameters. 
\lpsep is located at a distance of $14.6 \pm 0.1$\,pc \citep{Reid02.1}. 
\citet{Mohanty03.1} derive an effective temperature of $2550$\,K, and from the $J$ band magnitude
we obtain $L_{\rm bol} = 2.09 \cdot 10^{30}$\,${\rm erg\,s^{-1}}$ using the bolometric correction 
$(B.C.)_{\rm J} = 1.95$ \citep{Reid01.1}. 
According to the evolutionary models of \citet{Chabrier00.2} 
\lpsep is close to the bottom of the main-sequence ($0.04-0.1\,M_\odot$) and
moderately young ($\approx 0.3$\,Gyr).   
\lpsep belongs to the strongest H$\alpha$ emitters of its spectral type.   
Its H$\alpha$ equivalent width has been measured at various epochs
with values between $20...80\,\AA$,  
\citep{Martin96.1, Gizis00.1, Reid02.2, Mohanty03.1}. 
The magnetic field strength of \lpsep ($>3.9$\,kG) is 
among the highest in a sample of ultracool dwarfs studied by \citet{Reiners06.1}. 
No reports on X-ray emission from \lpsep exist in the literature, 
and in particular the {\em ROSAT} All-Sky Survey data implies 
a flux limit of $<2 \times 10^{-13}\,{\rm erg\,cm^{-2}\,s^{-1}}$, corresponding to a 
soft X-ray luminosity of $<5 \cdot 10^{27}$\,${\rm erg\,s^{-1}}$.

\section{Observations and data analysis}\label{sect:obs_and_anal}

We observed \lpsep for $40$\,ksec with 
{\em XMM-Newton}\footnote{The satellite and its instruments have been described 
in a Special Issue of Astronomy \& Astrophysics Vol. 365 (Jan. 2001)} 
(Obs-ID 0300170101), and analysed the data 
with the {\em XMM-Newton} Science Analysis system (SAS) version 6.5.0. 

The thin filter was inserted for all three European Photon Imaging Camera (EPIC) 
instruments. In this paper we use only data from the EPIC/pn CCD. Data processing included 
standard filtering of the event list.
After removing time intervals of high background using a screening algorithm,  
the EPIC/pn yielded $25$\,ksec of useful science data.
Visual inspection of the X-ray image showed an unexpectedly bright X-ray source 
coincident with \lp. For the temporal and spectral analysis of this source we 
used the source photons extracted from 
a $30^{\prime\prime}$ circle centered on the X-ray position of \lpsep 
and background photons from a circle of equal size positioned on a source-free
adjacent position. 

The Optical Monitor (OM) onboard  {\em XMM-Newton} was operated in the standard {\sc Image Fast} Mode 
with the $V$ band filter. 
For this work we use only data from the central fast-mode window of $10.5^{\prime\prime} \times 10.5^{\prime\prime}$ pixels 
read out with a time resolution $20$\,s. 
For technical reasons the OM observation had to be divided into nine exposures 
($7 \times 4400$\,s, $1 \times 3700$\,s, and $1 \times 2960$\,s) separated by 
an overhead of $\sim 0.3$\,ksec. Thus, there is almost complete time coverage
of the X-ray observation in the $V$ band with high time resolution.  
The OM data from the central fast-mode window was processed with 
standard SAS procedures, 
resulting in a sequence of $V$ band count rates with time resolution of $20$\,s.  
For the conversion of OM count rates into magnitudes and fluxes we
used the numbers given in the {\em XMM-Newton} SAS Users' 
Guide\footnote{http://xmm.vilspa.esa.es/external/xmm$\_$user$\_$support/documentation /uhb/index.html}.

\begin{figure}
\begin{center}
\resizebox{9cm}{!}{\includegraphics{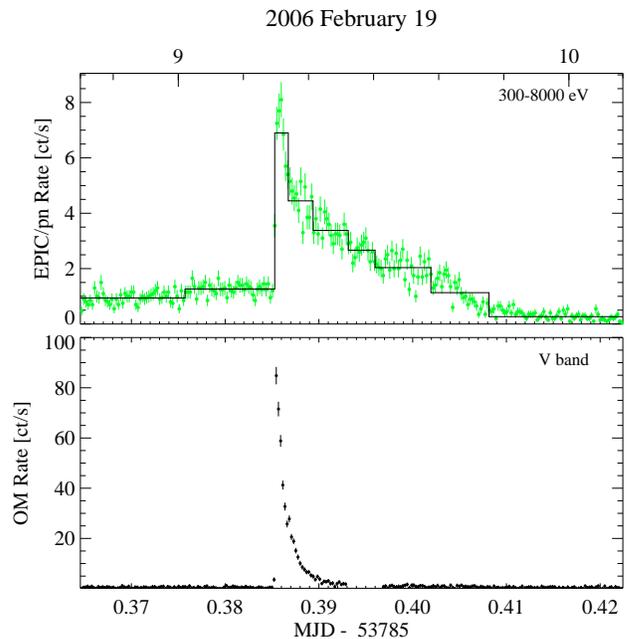}}
\caption{Fraction of the {\em XMM-Newton} lightcurves of \lpsep showing the giant flare and the precursor state. 
Both EPIC/pn and OM data are binned to the $20$\,s time resolution given by the OM.  
The EPIC/pn lightcurve is overlaid by the blocks derived in the ML analysis. 
The last block is not seen because of the restricted x-scale.}
\label{fig:lcs_part}
\end{center}
\vspace{-0.75cm}
\end{figure}

An inspection of the lightcurves of both the X-ray and optical data showed the presence of a huge
flare at about one third into the observation, 
but a weak signal without any signs of variability during the rest of the observation. 
Substantial precursor activity 
before the actual flare onset is clearly visible in the X-ray data (EPIC/pn count rate enhanced by 
a factor $\approx 30$ with respect to the pre- and post-flare quiescent emission). 
Fig.~\ref{fig:lcs_part} shows a fraction of the recorded lightcurves, starting with 
the X-ray flare precursor, during which no significant activity is seen in the OM data. 
In Table~\ref{tab:flare} we summarize count rates, magnitudes, and luminosities measured during the
different observed activity states. 

For the quiescent state before and after the flare (not shown in Fig.~\ref{fig:lcs_part}), we measure an  
X-ray luminosity of $\log{L_{\rm X,q}} \sim 27.2$\,${\rm erg\,s^{-1}}$,
corresponding to a fractional X-ray luminosity of $\log{(L_{\rm X,q}/L_{\rm bol})} \sim -3.1$.  
This represents the first detection of (quiescent) X-ray emission for \lp. Our measurement  
puts \lpsep at the upper end of the activity range for flare stars, not unexpectedly 
considering its strong and variable H$\alpha$ emission and strong magnetic field.  
The quiescent OM $V$ band count rate translates into a $V$ band 
magnitude of $\sim 19$\,mag, in good agreement with the literature value \citep{Dahn02.1}. 
 
Both visual inspection of the lightcurves and Spearman's and Kendall's rank correlation tests
are consistent with a time delay of $\leq 20-40$\,s ($1-2$ bins) between the optical and X-ray bands.
Both EPIC/pn and OM count rates increase by factors of $200-300$ from quiescence to their peak luminosities
of $4.7 \cdot 10^{29}$\,${\rm erg\,s^{-1}}$ and $2.8 \cdot 10^{30}$\,${\rm erg\,s^{-1}}$, respectively. 
Specifically, the $V$ magnitude rises within 
$40$\,s to a maximum of $13.1$\,mag ($\Delta V \approx 6$\,mag), whence it
decays exponentially back to quiescence with 
decay time scale $\tau_{\rm OM} \sim 90$\,s.  The decay of the X-ray light curve is far more prolonged. 
It is fast initially, but slows down soon after and
can be roughly described by two exponentials, a first one with  
a decay time scale similar to the optical decay ($\tau_{\rm 1,pn} \sim 150$\,s), 
and a second exponential that decays with $\tau_{\rm 2,pn} \sim 1000$\,s. 

%
%

We analysed the EPIC/pn light curve with a binning independent maximum likelihood (ML) method
described in detail by \citet{Stelzer06.1}.
We use this technique first to find possible background variations and then subtract the background
from the events listed at the source position, making use of 
the previously determined segmentation of the background signal. The result is
a background subtracted net source photon events list, to which the ML algorithm is applied.
In Fig.~\ref{fig:lcs_part} the binned lightcurve is overlaid by 
the constant rate blocks resulting from the ML analysis with an
imposed minimum number of $500$\,cts per block and a confidence level for change points of $99.9$\,\%. 
Since the data before the flare precursor (not shown in Fig.~\ref{fig:lcs_part}) 
consists of fewer than $500$\,cts, 
we start the ML analysis with the flare precursor. 
This procedure results in a division of the lightcurve into ten intervals of constant signal which we
use for spectral analysis.
%
%
\begin{table}\tiny
\begin{center}
\caption{\normalsize Count rates, magnitudes and luminosities for \lpsep in different activity states.} 
\label{tab:flare}
\begin{tabular}{llrrr} 
\hline 
               & & \multicolumn{1}{c}{pre-flare$^{(a)}$}  & \multicolumn{1}{c}{flare peak} & \multicolumn{1}{c}{post-flare$^{(b)}$} \\
\hline
\multicolumn{5}{c}{OM} \\
\hline
 $C$  & [${\rm ct\,s^{-1}}$]         &  $0.50 \pm 0.20$      & $84.83 \pm 3.47$    & $0.30 \pm 0.08$      \\
 $V$  & [mag]                        &  $18.7 \pm 0.4$       & $13.1 \pm 0.1$      & $19.3 \pm 0.2$       \\
 $L_{\rm V}$ & [${\rm erg\,s^{-1}}$] &  $(1.6 \pm 0.5) \cdot 10^{28}$  & $(2.8 \pm 0.1) \cdot 10^{30}$ & $(9.8 \pm 2.0) \cdot 10^{27}$  \\
\hline
\multicolumn{5}{c}{EPIC/pn} \\
\hline
 $C$  & [${\rm ct\,s^{-1}}$]         &  $1.089 \pm 0.067$    & $8.100 \pm 0.636$   & $0.042 \pm 0.003$    \\
 $L_{\rm X}$ & [${\rm erg\,s^{-1}}$] &  $(6.1 \pm 0.4) \cdot 10^{28}$  & $(4.6 \pm 0.4) \cdot 10^{29}$ & $(1.6 \pm 0.2) \cdot 10^{27}$  \\
\hline
\multicolumn{5}{l}{$^{(a)}$ defined by ML blocks No.1 and~2 representing the flare precursor} \\
\multicolumn{5}{l}{ activity (enhanced count rate in EPIC/pn with respect to quiescence; see text).} \\
\multicolumn{5}{l}{$^{(b)}$ defined by ML block No.10 comprising all data after the flare decay.} \\
\end{tabular}
\end{center}
\vspace{-0.5cm}
\end{table}

For each of the ten blocks as well as for the first $6$\,ksec of the observation
(representing the quiescent emission before the flare)   
a source and background spectrum, a response matrix and an 
ancilliary response file were extracted with standard {\em XMM-Newton} SAS tools.
We first fitted the quiescent spectrum within the XSPEC v11.3 environment using
a two-component {\sc APEC} thermal spectrum with a photo-absorption term. 
The spectral fitting shows the absorbing column to be very low as expected for a nearby star.
For the subsequent analysis of the flare spectra we froze all parameters of the 
quiescent model and added a spectral model describing the additional flare emission.

\begin{figure}
\begin{center}
\resizebox{7cm}{!}{\includegraphics{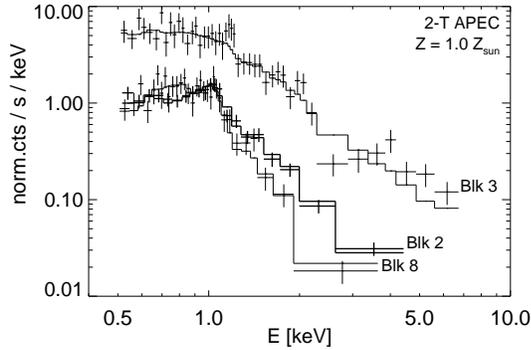}}
\caption{EPIC/pn spectrum and best fit model for flare precursor (Blk\,2; thick lines), flare peak (Blk\,3) and the end of the flare decay (Blk\,8).}
\label{fig:spec}
\end{center}
\vspace{-0.75cm}
\end{figure}

\begin{figure}
\begin{center}
\resizebox{7cm}{!}{\includegraphics{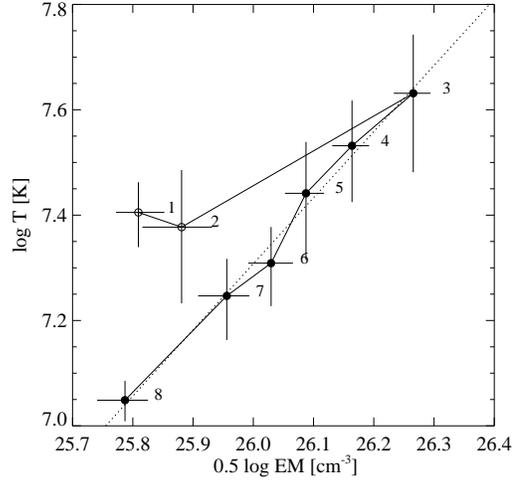}}
\caption{Time evolution of coronal temperature and emission measure. Open circles denote the flare precursor,
filled circles the flare decay. The dotted line represents a linear least squares fit to the flare
decay, that yields a slope $\zeta = 1.25$. 
Numbers in the plot refer to the constant rate blocks determined with the ML analysis.} 
\label{fig:t_em}
\end{center}
\vspace{-0.75cm}
\end{figure}

For the flare component, 
we tested a series of one-temperature ($1$-T) and two-temperature ($2$-T) thermal models with
global abundance $Z$ set to different values 
($Z = 0.3\,Z_\odot$, $Z = 1.0\,Z_\odot$, and $Z$ a free fit variable). 
The $1$-T models turn out to be inacceptable for most time segments.
The only approach that yields 
a statistically acceptable description and well-constrained parameters 
for all spectra is the $2$-T model with $Z = 1.0\,Z_\odot$, considered
the best fit model (see Fig.~\ref{fig:spec} for a selection of spectra). 
%
%
A mean temperature $T$ was computed for each spectrum from an emission measure ($EM$) weighted sum of the 
two fit components.  
The total $EM$ is given by the sum of the $EM$s of both components. 
The resulting $T$ vs. $\sqrt{EM}$ diagram is plotted in Fig.~\ref{fig:t_em} and will be used as
a diagnostic for the flare evolution.

\section{Discussion and Conclusions}

\subsection{Optical and X-ray properties of the giant flare}

To derive quantitative information from the optical signal 
we choose a description of the flare photosphere in 
terms of blackbody emission with effective temperature as a free parameter. 
Although the flaring photosphere is likely not a perfect blackbody, 
at least in the $V$ band there are no strong emission lines \citep{deJager89.1}, 
and blackbody models have provided a reasonable approximation for 
flare continua on earlier M dwarfs \citep{Hawley92.1}. 
Using the $V$ filter response curve, 
the luminosity derived from the OM count rate (see Table~\ref{tab:flare}) 
can be corrected for the temperature dependent out-of-band flux ("bolometric correction").
Assuming a photospheric `flare temperature' $T_{\rm eff} = 9500$\,K, 
we obtain an optical flare peak luminosity of 
$L_{\rm bol, peak} = 4.7 \cdot 10^{30}$\,${\rm erg\,s^{-1}}$; 
raising $T_{\rm eff}$ to $15000$\,K increases this luminosity by a factor of two.
With an assumed stellar radius of $0.12\,R_{\odot}$, the area
of the flaring region on the star is $(3.4 - 13)\,10^{18}\,{\rm cm^{2}}$,
i.e., the flare occupied $< 1$\,\% of the surface of \lp.  
This compares well with the flare area sizes determined for UV Cet \citep{deJager89.1}
and for solar white light flares \citep{Xu06.1}.

%
%
We determined the physical parameters of the X-ray emitting region making use of the hydrodynamic flare model 
and its calibration to the EPIC/pn spectral response described by \citet{Reale04.1}. This approach implies
that we interpret the flare on \lpsep as a compact loop flare.
From the measured slope of $\zeta = 1.25$ of the flare evolution in the $\log{T} - \log{\sqrt{EM}}$
plane (see Fig.~\ref{fig:t_em}) and the observed $\log T_{\rm peak}\,{\rm [K]} = 7.65$, 
we find a (half) loop length of $L \approx 1.4 \cdot 10^{10}$\,cm, i.e.,
a value larger than the radius of \lp.  Assuming that the cross section of the flaring loop be given by 
the size of the optical flaring region, we compute a volume of 
$V_{\rm flare} \approx (0.9-3.6) \cdot 10^{29}\,{\rm cm^{3}}$ and 
-- from the observed peak $EM$ of $3.4 \cdot 10^{52}\,{\rm cm^{-3}}$ -- a mean plasma density of 
$(3-6) \cdot 10^{11}\,{\rm cm^{-3}}$. 
In principle, the expectation of high density could be checked with the {\em XMM-Newton} RGS data,
but unfortunately due to the high flare plasma temperature 
the signal in the O~VII triplet (which would be sensitive to densities in that range)
is so weak that no quantitative conclusions can be drawn from that data. At any rate, however, 
the size of the flaring region {\it in X-rays} is quite large and of the order of the stellar
dimensions.  Therefore considerable heating also of the non-flaring atmosphere must occur as a result
of its illumination by the hot flaring plasma. Further, the derived length scale $L$ must be filled
with plasma quite rapidly.  Assuming {\em ad hoc} expansion velocities of $1000$\,km/s, the typical
filling time scales are of the order of $50  - 100$ \,s, which is comparable to the possible observed delay 
between flare onset and flare maximum.  Also, considerable line shifts and broadening are expected
which, however, cannot be diagnosed because of low signal-to-noise of the presently available data.  
Finally, we can estimate the total thermal energy content 
$E_{\rm th,X} =  3 n k T V$  of the X-ray plasma at flare peak. Using the above numbers we find 
$E_{\rm th,X} =  (1 - 2) \cdot 10^{33}$\,erg.

In terms of optical amplitude this flare on \lpsep is among
the largest recorded in detail throughout the history of stellar flare observations and
it belongs to the very few flares allowing an assessment of the overall energy budget 
both in X-ray and optical light. The emitted flare energy amounts to $3 \cdot 10^{32}$\,erg,
each in the $V$ band and in soft X-rays (see Table~\ref{tab:flare}), values
clearly at the upper
end of the range observed for field flare stars but lower than those for T Tauri stars 
\citep{Fernandez04.1, Stelzer06.1}.  The total radiated energy is thus at least
$E_{\rm rad} \approx 6 \cdot 10^{32}$\,erg, and this is obviously a lower limit, since
energy may be radiated in other wavebands.  
We further find that the energy radiated in the soft X-ray band 
is roughly of the same order as the value derived for the thermal energy content 
of the coronal plasma at flare peak,  
and hence there appears to be no substantial energy release required in the flare decay phase. 
This is consistent with the fact that the observed value for the slope in the $T$ vs. $\sqrt{EM}$ diagram 
is close to the limiting value of $\zeta = 1.3$, above which the heating time scale is zero
in the hydrodynamic model of \citet{Reale04.1}. 

In the standard flare scenario it is assumed that 
the total energy released in the flare is ultimately derived from energy stored in 
the magnetic field and converted into other forms of energy during the flare. 
Comparing the magnetic energy density to the radiated flare energy and flare volume derived above, 
we can compute the mean magnetic field that must be annihilated in the corona 
in order to provide the observed energy. We find 
$B_{\rm annih} \approx (140 - 280)$\,G. 
The minimal field necessary for confinement is the equipartition field,  
$B_{\rm eq} \approx 310 - 430$\,G,  
and obviously this field must not be destroyed in the course of the flare
progess. The total coronal magnetic field prior to flare onset must thus be in the
range $450-710$\,G in a volume of 
$V_{\rm flare} \approx (9-36) \cdot 10^{28}\,{\rm cm^{3}}$.  
If the annihilated field is limited to a smaller volume, the field strength must
be even larger. The flare onset is quite rapid and the conversion of
magnetic energy into thermal energy occurs within $20 - 40$\,s at most.

\subsection{\lpsep in the context of stellar flares}

The so far most detailed example of a coordinated optical and X-ray observation 
of a large flare on an M dwarf presented in the literature is an {\em XMM-Newton} 
observation of Proxima\,Cen (M5.5e) by \citet{Guedel04.1}.  The peak X-ray luminosities of
the flares on Proxima\,Cen and \lpsep are quite similar and it is instructive to
compare them in more detail. As in \lp, the optical flare on Proxima\,Cen was shorter than the X-ray event, 
in line with the expectation from a flare scenario where the optical burst is considered as a proxy of the
initial (impulsive) heating of chromospheric plasma by accelerated electrons.
However, there are marked phenomenological differences between
the two events: 
The Proxima\,Cen X-ray flare was a long-duration event with a rise time of $12$\,min and decay 
time of a significant fraction of a day, 
in contrast to the short flare observed on \lp. 
The onset of the optical event preceeded that of the X-ray flare in the case of Proxima\,Cen 
while in \lpsep any time lags are below the time resolution of $20$\,s. 
In the chromospheric evaporation scenario stellar flares are expected to peak in the 
optical before the X-rays, but due to the lack of simultaneous multiband data 
observational evidence was found only in few cases \citep{Schmitt93.1, MitraKraev05.1}. 

The EPIC light curve of Proxima\,Cen displays -- unlike the case of \lpsep -- 
next to a major flare with total energy $\approx 1.5 \cdot 10^{32}$\,erg, 
several small events with energies of a few $10^{29}$ to $\approx 10^{31}$\,erg.
Such observations of lower level flaring have lead to the
idea of stellar coronal nano-flare heating. 
Studies of solar flares and some bright flare stars suggest the energy distribution of flares
to follow a power law with slope $\approx -2$ in its differential form 
\citep[e.g.,][]{Aschwanden00.1, Audard00.1}. 
A systematic analysis of X-ray flaring in a sample of pre-main sequence stars in the Taurus star 
forming region has yielded similar results \citep{Stelzer06.1}.   
In the context of \lpsep arises the question if the observed giant flare can be
interpreted as the "high energy tail" of a canonical flare frequency distribution.
According to \citet{Stelzer06.1} we expect to be sensitive to X-ray events with 
relative amplitude $A_{\rm rel} = (R_{\rm max} - R_{\rm Q})/R_{\rm Q} > 0.3$,
where $R_{\rm max}$ and $R_{\rm Q}$ are the maximum and the quiescent soft X-ray count rate, 
given the sensitivity limit imposed by Poisson statistics and the observed quiescent EPIC/pn count rate of \lp.
This limit corresponds to flares with peak luminosity 
of $\log{L_{\rm X}}\,{\rm [{\rm erg\,s^{-1}}]} \sim 27.3$. Assuming a duration of $5000$\,s for such 
micro-flares -- probably a conservative estimate considering the
short duration of the observed giant flare -- the emitted energy of such events
would be $3 \cdot 10^{30}$\,erg. If the number distribution of flares on \lpsep were 
as observed for other stars, one would expect to see $\approx 100$ such events for each giant flare, 
in clear contrast to our EPIC/pn data. Depending on the absolute flare frequency, it would in principle 
be possible that we failed to detect smaller flares because of the limited observing time, 
however, it is as well possible that the actual flare energy distribution of \lpsep is rather flat.
In this case, the origin of the quiescent X-ray emission 
(often considered to be composed of unresolved micro- and nano-flares) remains unclear. 
Future multi-wavelength monitoring observations of ultracool dwarfs will prove useful to 
constrain the structure and energetics of their outer atmospheres. 

\begin{acknowledgements}
We thank the referee C. Bailer-Jones for his quick response. 
BS and GM acknowledge funding by ASI-INAF contract I/023/05/0. CL was supported by DLR grant OR5010.
This work is based on observations obtained with {\em XMM-Newton}, an ESA science mission with instruments and 
contributions directly funded by ESA Member States and NASA. 
\end{acknowledgements}
\vspace{-0.5cm}

\bibliographystyle{aa}
\bibliography{aa6488}

\end{document}